\documentclass[apj]{emulateapj}
\usepackage{apjfonts}

  \usepackage{amscd}
  \usepackage{amsmath}
  \usepackage{amssymb}
  \usepackage{verbatim}
  \usepackage{array}
  \usepackage{psfrag}

  \shorttitle{3D deflagration SNe Ia model}
  \shortauthors{F.~K.~R{\"o}pke et al.}

\begin{document}

\title{A three-dimensional deflagration model for Type Ia supernovae
  confronted with observations}

\author{F. K. R{\"o}pke\altaffilmark{1,2,3},  W. Hillebrandt\altaffilmark{1,3},
    W. Schmidt\altaffilmark{4}, J. C. Niemeyer\altaffilmark{4}, S. I.
    Blinnikov\altaffilmark{1,5}, and
    P. A. Mazzali\altaffilmark{1,3,6}}
\altaffiltext{1}{Max-Planck-Institut f\"ur Astrophysik,
              Karl-Schwarzschild-Str. 1, D-85741 Garching, Germany}
\altaffiltext{2}{Department of Astronomy and Astrophysics, University
  of California Santa Cruz, 1156 High Street, Santa Cruz, CA 95064,
  U.S.A.}
\altaffiltext{3}{Kavli Institute for Theoretical Physics, University of
  California, Santa Barbara, CA 93106}
\altaffiltext{4}{Lehrstuhl f{\"u}r Astronomie und Astrophysik,
  Universit{\"a}t W{\"u}rzburg, Am Hubland, D-97074 W{\"u}rzburg, Germany}
\altaffiltext{5}{SSC RF ITEP, Bolshaya Cheremushkinskaya 25, 117218
  Moscow, Russia}
\altaffiltext{6}{Istituto Nazionale di Astrofisica, OATs, via Tiepolo,
11, 34131 Trieste, Italy}

\begin{abstract}
A simulation of the thermonuclear explosion of a Chandrasekhar-mass C+O white dwarf, 
the most popular scenario of a type Ia supernova (SN~Ia), is presented. 
The underlying modeling is pursued in a self-consistent way, treating the combustion wave as 
a turbulent deflagration using well tested methods developed for laboratory 
combustion and based on the concept of `large eddy simulations' (LES).
Such consistency
requires to capture the onset of the turbulent cascade on resolved
scales.
This is achieved by computing the dynamical evolution 
on a 1024$^3$ moving grid, which resulted in the best-resolved
three-dimensional SN~Ia simulation 
carried out thus far, reaching the limits of what can be done on present 
supercomputers.
Consequently, the model has no free parameters other than the
initial conditions at the onset
of the explosion, and therefore it has considerable predictive power. 
Our main 
objective is to determine to which extent such a simulation can account 
for the observations of normal SNe~Ia. Guided by previous simulations with less 
resolution and a less sophisticated flame model, initial conditions were chosen 
that yield a reasonably strong explosion and a sufficient amount of radioactive
nickel for a bright display.  
We show that observables are indeed matched to a reasonable degree. 
In particular, good agreement is found with the light curves of normal SNe Ia. 
Moreover, the model reproduces the general features of the abundance
stratification as inferred from the analysis of spectra. This indicates that 
it captures the main features of the explosion mechanism of SNe Ia.
However, we also show that even a seemingly best-choice pure deflagration model 
has shortcomings that indicate the need for a different mode of nuclear burning
at late times, perhaps the transition to a detonation at low density. 
\end{abstract}


\maketitle

\section{Introduction}
\label{sect:intro}

Understanding the physics of Type Ia supernova (SN Ia) explosions is a necessary
premise for using them as cosmic distance indicators with
confidence. Consequently, the development of three-dimensional explosion models 
has been a major challenge in astrophysics over the past decade. Only 
recently, however, a self-consistent three-dimensional treatment of
the turbulent thermonuclear flame in a white dwarf (WD) star avoiding
free parameters became available.
The ultimate goal of 
these efforts is to describe the full observational set of SNe Ia and even to 
predict their observed properties, such as their luminosities, light curves,
and spectra, from first principles, with sufficient accuracy to guide the 
empirical calibration of their peak luminosity. 

While this stage of sophistication is clearly not yet reached with current
simulations, at least one (perhaps in some cases the dominant) part of the mechanism can already
be treated in an non-parameterized approach. In the first stage of the supernova
explosion the flame propagates subsonically and is accelerated by turbulence
creating a problem of turbulent combustion physics. The knowledge on this
subject acquired for laboratory combustion in engineering sciences can be
transferred to astrophysics, facilitating self-consistent simulations
with particular emphasis on the correct representation of the flame/turbulence interaction
\citep{niemeyer1995b,reinecke1999a,reinecke1999b,reinecke2002d,reinecke2002b,roepke2005c,roepke2005b,roepke2006a,schmidt2006a,schmidt2006c}.
In fact, most of the properties of turbulent chemical flames 
are recovered in SNe~Ia, and the largely different relevant length scales,
the simpler reaction kinetics, and the absence of complicated boundaries, make 
computations of thermonuclear supernovae perhaps easier than designing a new
car engine, provided the initial conditions of the explosion are known. 

Recent results from several three-dimensional simulations
\citep{reinecke2002d,gamezo2003a,travaglio2004a,roepke2005b,bravo2006a,roepke2006a,schmidt2006a}
indicate that a turbulent thermonuclear deflagration---a subsonic
flame propagation---alone can
lead to powerful explosions of the WD star. 
In combination with the elimination of free parameters in the model 
the question arises to what extent the results of such simulations match observational
constraints. This question motivated several previous studies which
addressed initial parameters of the exploding WD
\citep{roepke2004c,travaglio2005a,roepke2006b}, the flame ignition
configuration \citep{reinecke2002d,roepke2006a,schmidt2006a}, symmetry constraints
in the simulation setup \citep{roepke2005b}, and the particular choice of the turbulence
model \citep{schmidt2006c}, as well as the impact of some initial condiditions
on observational results \citep{blinnikov2006a}. 

One question left out until now, however, was whether the results would
change significantly with higher resolution. The flame/turbulence
interaction determines the propagation of the burning front and thus
the results of the model with respect to nucleosynthesis and
energetics. Given the expected strong turbulence effects and the
associated vast range in relevant length scales this endeavor may
seem hopeless. However, as discussed below, with the numerical
modeling strategy pursued here, it turns out to be sufficient to
resolve the onset of turbulence on the computational grid in order to
recover a definitive result. The necessary resolution can be
reached with current computational resources and
marks the aim of the present study.
A direct comparison with observational data is possible and
is indeed pursued in this publication for one particular simulation in
the framework of the pure deflagration model. 
With what is so far the best 
resolved three-dimensional simulation of the deflagration scenario for SNe Ia
we assess its consistency and its compatibility with observations. 
As we show in this paper, such a state-of-the-art deflagration model
matches observational data surprisingly well, although not all details
are reproduced perfectly. As is argued below, the simulation reaches a state 
where the physical effects are resolved so that its predictions are
expected to be robust and to reflect generic properties of the deflagration
scenario.

Of course, a single simulation cannot provide the basis for general
conclusions on the validity of the deflagration model. Here, we
present a simulation performed within the range of realistic
initial conditions (as far as known) but explicitly set up the initial flame
configuration in a favorable way with regard to energy and
$^{56}$Ni production.
Combining this single realization of the deflagration model with the
exploration of the parameter space carried out
previously (and partially with simpler modeling approaches and less resolved
simulations), successes and limitations of the deflagration model
become visible and will be discussed accordingly (see Sect.~\ref{sect:limits}).    
In particular, some persisting shortcoming of the deflagration model
can be taken as an indication that the deflagration stage alone cannot
explain all aspects of SNe Ia.
In some cases models beyond the deflagration regime are supported by observational results 
\citep[e.g.][]{mazzali2007a,wang2007a}, but uncertain or even hypothetical 
physical mechanisms generally introduce free parameters into the models and 
make a face-to-face comparison with observations more difficult than
in the case of the pure deflagration model.

\section{Astrophysical model}
\label{sect:model}

Recent observations \citep[e.g.][]{mazzali2007a} support the scenario that the
majority of SNe~Ia originate from a carbon-oxygen WD disrupted by a
thermonuclear explosion \citep{hoyle1960a} when it approaches the
Chandrasekhar mass limit by 
mass accretion from a binary companion which may perhaps be a
non-degenerate star (\citealp{ruiz-lapuente2004a}; for alternative
progenitor scenarios see e.g.\ \citealp{hillebrandt2000a}). Caused by 
temperature fluctuations inside the convective core of the electron-degenerate
WD, small regions near its center are thought to commence a thermonuclear
runaway with rapidly increasing nuclear energy release from burning the
carbon-oxygen fuel to heavier elements. This process marks the birth of the
thermonuclear flame. The exact number and distribution of ignition kernels is
not well determined yet, implying that the largest uncertainty lies in the 
initial parameters of the explosion. The virtue of three-dimensional 
explosion models is that they directly relate the ignition conditions to the
outcome of the explosion without introducing additional
parameters. Thus a comparison with observations
sheds light on two aspects of the supernova problem---the progenitor evolution
giving rise to these initial conditions and the explosion mechanism itself. 

Here, we are concerned with the latter, for which several possibilities have 
been suggested. Hydrodynamically, two fundamentally distinct modes of flame
propagation are admissible: a subsonic deflagration and a supersonic
detonation. A prompt detonation would incinerate the entire star at the high 
initial densities, leading to conditions where nuclear reactions produce almost 
exclusively iron group elements \citep{arnett1971a}. These are found to be 
abundant in the ejecta, predominantly produced as radioactive $^{56}$Ni. 
In its decay to $^{56}$Co (and later $^{56}$Fe) gamma-rays and positrons are 
emitted which down-scatter to optical wavelengths in the SN ejecta and 
give rise to the observed optical display. At later times, in the
so-called ``nebular'' spectra, this iron can be directly observed. However, 
failure to produce the intermediate mass elements (e.g.\ silicon, calcium, and 
sulfur) observed with appreciable abundances in the early spectra of SNe~Ia 
\citep[for a review see][]{filippenko1997a} rules out the prompt
detonation mechanism. 
In contrast, a subsonic deflagration leaves enough time for the stellar material
to expand before it is reached by the thermonuclear flame
\citep{nomoto1976a}. Therefore, some
burning takes place at densities low enough such that mostly intermediate mass
elements are synthesized, while most $^{56}$Ni is produced during
the early deflagration phase.

A laminar
deflagration wave, mediated by microphysical heat transport, would propagate
far too slowly to explode the WD. However, as burning ignites near the
WD's center and propagates outward, a stratification of low-density ashes  
below denser fuel emerges which is
buoyancy-unstable in the gravitational field of the star. This
configuration eventually leads to the formation of ascending burning
bubbles. Their interfaces are subject to shear. Since typical Reynolds
numbers of the flow are of the order of 10$^{14}$, strong turbulent eddies must
form which decay to smaller scales in a turbulent cascade.
This has tremendous impact on the flame propagation. The flame surface is 
corrugated by turbulent velocity
fluctuations and its area increases dramatically, which enhances the 
burning rate. This acceleration of the flame speed and the 
fuel consumption rate enable
explosions of the WD \citep[e.g.][]{reinecke2002d}.

The energy input into turbulent eddies is due to the
Rayleigh-Taylor instability on the largest scales
($\sim$$100\,\mathrm{km}$). However, the flame 
interacts with turbulent velocity fluctuations down to much smaller
scales. As demonstrated in Sect.~\ref{sect:turb} below, these velocity
fluctuations originate from an inertia-driven turbulent cascade. They
do not obey a simple buoyancy-dominated scaling and therefore require
special modeling effort.

\section{Numerical implementation}
\label{sect:num}

Numerical schemes to model the propagation of a deflagration front in a SN~Ia
have been developed over the past years and have reached a high level of
sophistication. 
The major challenge in the numerical implementation of the concept
outlined in Sect.~\ref{sect:model} is posed by the vast
range of relevant scales. It spans about
11 orders of magnitude, from the radius of the WD ($\sim 2000 \,
\mathrm{km}$ and expanding in the explosion) down to the Kolmogorov scale
(less than a millimeter) where turbulent energy is
dissipated. The flame interacts with
turbulence down to the Gibson scale at which the laminar flame
propagation speed becomes comparable to the turbulent velocity
fluctuations. For most parts of the supernova explosion this scale is
considerably larger than the flame width ($\lesssim 1\,
\mathrm{mm}$). Therefore, the interaction of the flame with turbulence is
purely kinematic and burning takes place in the so-called \emph{flamelet
regime of turbulent combustion} \citep[e.g.][]{peters2000a}. Here, the flame propagation speed
completely decouples from the microphysics of the burning and is
entirely determined by turbulence effects. This also implies that the
flame propagation does not scale with the laminar burning speed.

The challenge of scales can be tackled in an elegant way by employing
a \emph{large eddy simulation} (LES) strategy. This approach resolves only
the largest scales of the
turbulent cascade and models the unresolved scales, so that it becomes
possible to accommodate the entire WD on the computational
domain. When seen
from the scale of the WD, the flame appears as a sharp discontinuity
separating the fuel from the ashes. A suitable numerical scheme to
track the propagation of such an interface is the level set method
\citep{osher1988a,smiljanovski1997a,reinecke1999a}, where the flame
front is associated with the zero level set of a signed 
distance function $G$ set up to be negative in the fuel regions and positive in
the ashes. The propagation velocity of this interface is
given by the effective turbulent burning speed at the grid scale which
in the flamelet regime is completely determined by turbulent velocity fluctuations on that
scale. These, in turn, are derived from a subgrid-scale turbulence model 
\citep[for details see][]{schmidt2006b,schmidt2006c} which follows the properties of
turbulence on
unresolved scales based on filtering the velocity fluctuations on the
resolved scales. This way, a localized treatment of the turbulent
properties of the flow becomes feasible and the flame/turbulence
interaction is modeled self-consistently. 

The hydrodynamics is described by the reactive Euler
equations (with gravity as external force) and
numerically---discretized on a Cartesian computational grid---treated
in a finite 
volume approach \citep{fryxell1989a}. Of course, the tremendous
expansion of the WD during the 
explosion has to be taken into account. This was achieved by employing
two nested moving grids---a fine uniform inner grid tracking the
flame and a coarse outer grid capturing the WD
\citep{roepke2005c,roepke2006a}.

With this implementation we performed what is thus far the largest simulation
of a thermonuclear supernova explosion. It was set up on more than a
billion computational grid cells (a $1024^3$ cells grid). This allowed for
a very fine resolution (down to $\sim 870\, \mathrm{m}$) of the central
part of the WD at the beginning of the explosion facilitating an
exceptionally 
fine-structured initial flame configuration that was composed of 1600
spherical kernels of radius $2.6 \, \mathrm{km}$ distributed inside a
sphere extending $180 \, \mathrm{km}$ around the center of the WD (see
Fig.~\ref{fig:evo}, upper left panel). Although
the details of the ignition process are yet unknown, analytical and
numerical studies support such a multi-spot ignition configuration
\citep[e.g.][]{garcia1995a,woosley2004a} and the flame ignition kernels are 
expected to be of the order of a kilometer in diameter \citep{iapichino2006a}.
The configuration chosen here is certainly one that
favors burning near the center and \citep[as multi-spot ignition scenarios
do in general,][]{roepke2006a} enhances the burning and energy release over that found
in centrally or sparsely ignited scenarios.

The nuclear reactions were treated in the approximative approach outlined by
\citet{reinecke2002b}. Only five species were taken into account: the
fuel consisted of a mixture of carbon and oxygen in equal amounts by mass,
intermediate mass elements were represented by magnesium, and a mixture
of $^{56}$Ni and alpha-particles represented nuclear statistical
equilibrium (NSE) material. Depending on the fuel density ahead of the
flame, $\rho_\mathrm{u}$, the material was converted to NSE (if
$\rho_\mathrm{u} > 5.25 \times 10^{7} \, \mathrm{g} \,
\mathrm{cm}^{-3}$), in which case the nickel/alpha ratio was adjusted
according to temperature and density, to intermediate mass elements (if
$5.25 \times 10^{7} \, \mathrm{g} \,
\mathrm{cm}^{-3} > \rho_\mathrm{u} > 5.0 \times 10^{5} \, \mathrm{g} \,
\mathrm{cm}^{-3}$), or left unburned (for lower fuel densities). The
threshold for burning to intermediate mass elements was chosen lower
than in previous simulations. As argued by \citet{schmidt2007a}, the
level-set description of the flame propagation may be extended beyond the
onset of distributed burning at which turbulent eddies penetrate the
internal structure (the ``thin reaction zones regime''). This
commences as soon as the flame width becomes
larger than the Gibson scale---an effect that is expected to occur at
fuel densities $\sim$$10^{7} \, \mathrm{g} \,
\mathrm{cm}^{-3}$. However, at densities below $\sim$$10^{6}
\, \mathrm{g} \, \mathrm{cm}^{-3}$ nuclear reactions will be slow
compared to the expansion time scales and, moreover, turbulence will
interact with the burning zone inside the flame structure giving rise
to the completely distributed (``broken reaction zones'')
regime. Therefore, the burning description adopted here may be too
optimistic at low densities and over-predict the total amount of
intermediate mass elements synthesized in the explosion. On the other
hand, at the
late stages where low fuel densities are reached, the amplitude of turbulent
velocity fluctuations decreases rapidly. Therefore not much material
is burned here, such that the fuel consumption at very low densities
becomes insignificant and the potential error introduced by
uncertainties is expected to be small. 

\section{Evolution of the explosion phase}
\label{sect:evo}

\begin{figure*}
\plotone{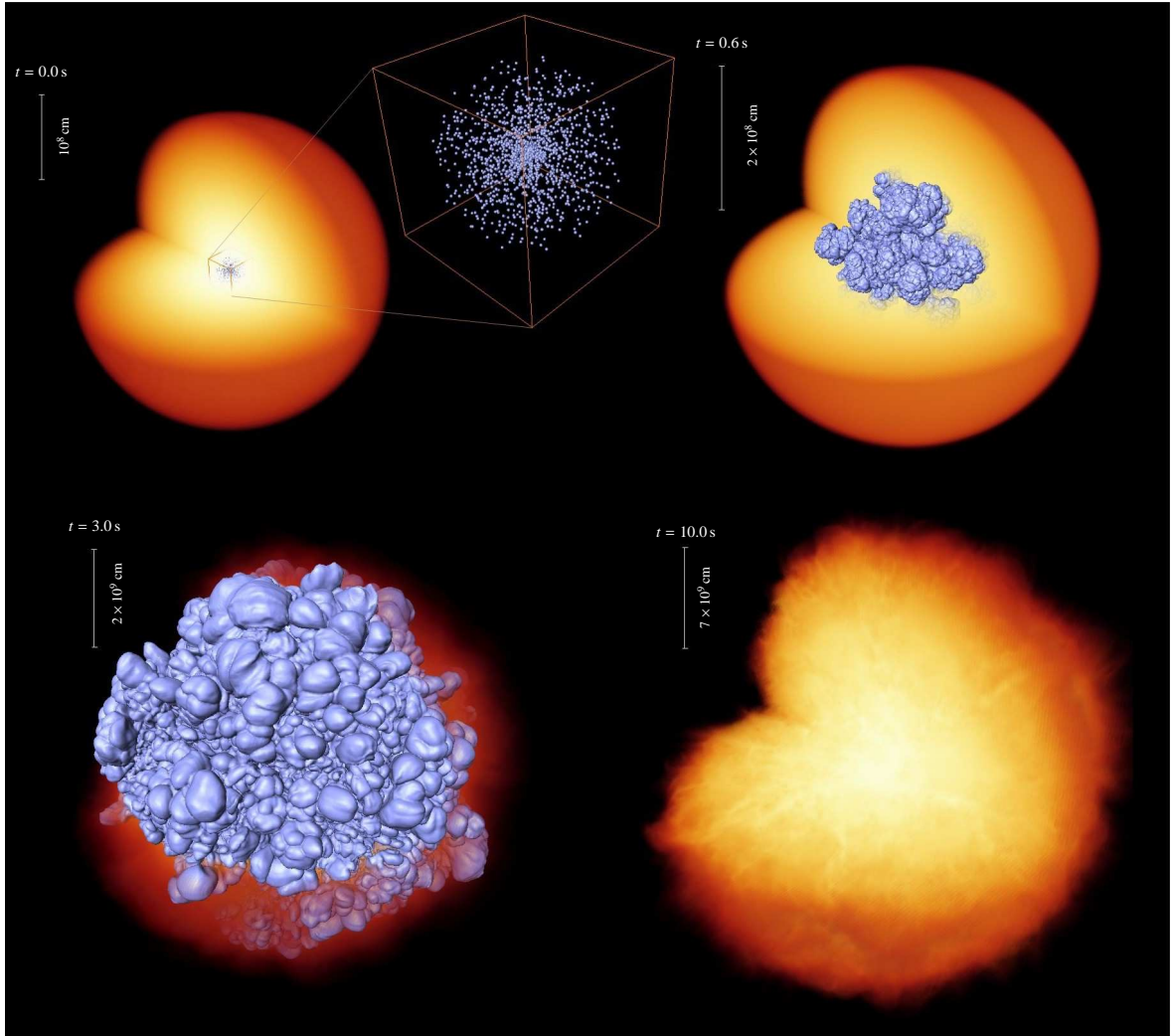}
\caption{Evolution of the thermonuclear supernova explosion
  simulation. The zero level set associated with the thermonuclear
  flame is shown as a blue isosurface and the extent of the WD  is
  indicated by the volume rendering of the density. The upper left
  panel shows the initial set up and the close-up illustrates the
  chosen flame ignition configuration. The subsequent two panels illustrate
  the propagation of the turbulent flame through the WD and the
  density structure of the remnant is shown in the lower right panel. \label{fig:evo}}
\end{figure*}

The evolution of the explosion stage proceeds in agreement with the
expectations outlined in Sect.~\ref{sect:intro}. After ignition in a
multitude of only partially overlapping flame kernels (see upper left
panel of Fig.~\ref{fig:evo}), the
ash region grows due to burning and starts floating towards the
surface. The individual kernels deform from their initially spherical
shape into ``mushroom cap''-like structures---typical for the
non-linear stage of the Rayleigh-Taylor instability---and grow. This causes 
them to merge, so that a 
connected flame structure forms. This structure evolves dominated by
buoyancy instability on large scales and driven by the turbulent
cascade on smaller lengths. The
continued development of substructure and the merger of features create a
deflagration structure with a complex pattern (see upper right panel of
Fig.~\ref{fig:evo}). Burning and flotation drive
the flame towards the surface of the WD. The fuel density drops as the
flame moves outwards and due to radial stratification and the overall expansion of the WD caused
by the energy deposit from nuclear burning. Once the fuel density
falls below the threshold for the production of intermediate mass
elements, nuclear burning ceases. This occurs first
at the leading features of the flame and subsequently in more central
flame regions. Finally, no burning takes place anymore. At this point
the outer ash features have reached the surface layers of the
ejecta (lower left panel of Fig.~\ref{fig:evo}). The following
seconds in the evolution are characterized by the hydrodynamical
relaxation towards homologous expansion. This stage is reached to good
approximation at about $10 \, \mathrm{s}$ after ignition \citep{roepke2005c}.

This three-dimensional evolution leads to a remnant of the explosion
with characteristic properties. The density structure has patterns from
unstable and turbulent flame propagation imprinted on it (cf.\ lower
right panel of Fig.~\ref{fig:evo}) and ash regions extend to 
the outermost layers of the expanding cloud of gas. 

\begin{figure*}
\plotone{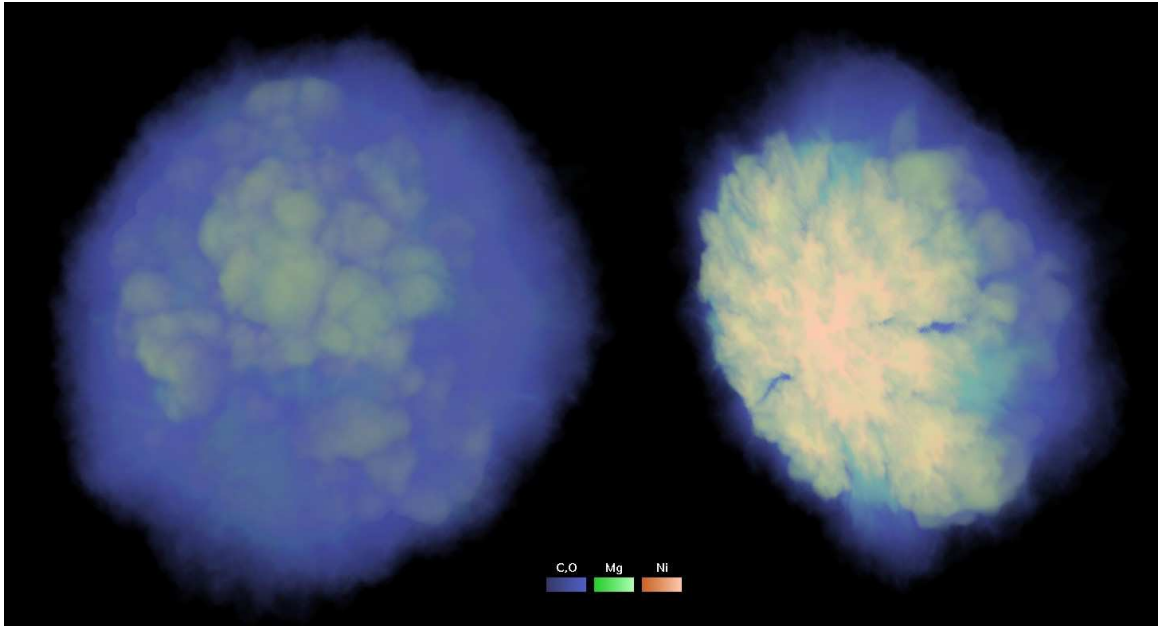}
\caption{Composition of the remnant after the explosion
  ($t=10\,\mathrm{s}$). The products of mass fractions of three of the species followed in the
  hydrodynamical explosion simulation with density are volume rendered in
  different colors. The right panel shows a cut-away to illustrate
  the composition of the central parts. \label{fig:comp3d}}
\end{figure*}

The chemical composition of the ejecta of the explosion as followed in the
hydrodynamical simulation is shown in Fig.~\ref{fig:comp3d}. The
outer parts consist mainly of unburned carbon and oxygen, but some
``bubbles'' of intermediate mass elements are also present near the
surface (Fig.~\ref{fig:comp3d} left). The central part of the ejecta
is dominated by iron group
elements (Fig.~\ref{fig:comp3d} right). Inside the ash bubbles some
mixing of different species is
found, but still regions dominated by either iron group elements or
intermediate mass elements can be identified. The large-scale bubbles
of ash are separated by a few thin but dense fingers of unburned
material so that the spherically averaged composition of the ejecta
appears strongly mixed.

\section{Consistency of the subgrid-scale modeling}
\label{sect:turb}

\begin{figure}
\plotone{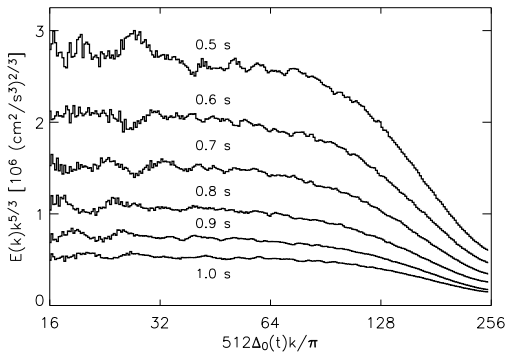}
\caption{Compensated energy spectrum functions derived from our
  simulation at different instants in time. \label{fig:turb}}
\end{figure}

The subgrid-scale model employed in the simulation couples the
dynamics at unresolved scales to the numerically computed flow. The
model is based on the balance equation for the subgrid-scale
turbulent energy, where the energy production rate follows from a
similarity assumption linking the smallest resolved to the largest
unresolved scales \citep{schmidt2006b}. Although the
calculation of the rate of turbulence energy production is localized
and, consequently, does not rely on stationarity or large-scale
homogeneity, asymptotic isotropy towards smaller length scales is
required for consistency. In the case of incompressible turbulence,
asymptotic isotropy becomes manifest in the Kolmogorov $k^{-5/3}$
scaling for sufficiently large wavenumbers $k$. This scaling has been
found in simulations of Rayleigh-Taylor unstable stratifications
\citep{cabot2006a} as well as in buoyancy unstable flames in
degenerate WD matter \citet{zingale2005a}.

Fig.~\ref{fig:turb} shows compensated energy spectrum functions,
$k^{5/3}E(k)$, derived from our simulation for several instants of time
ranging from $0.5$ to 
$1.0\,\mathrm{s}$ after ignition. For turbulent velocity fields
obeying Kolmogorov 
scaling, the compensated spectrum functions are expected to be
constant. Wavenumbers are normalized by $\pi/(512\Delta_{0})$, where
$\Delta_{0}$ is the cell size in the uniform part of the grid.  Since
the grid is inhomogeneous and does not satisfy periodic boundary
conditions, the velocity fields obtained from the simulation were
zero-padded in the outer regions with Gaussian windowing functions in
order to compute Fourier transforms. This is why the compensated
spectrum functions are plotted for normalized wavenumbers greater than
$16$ only. At all times considered here, it can be seen that the
spectrum functions are well approximated by the Kolmogorov power law
for intermediate wave numbers, thus indicating the presence of an
inertial subrange.  Therefore, the simulation is as well resolved as
the modeling approach requires. Higher resolution is not expected to
change the characteristics of the turbulent flame propagation.

\section{Results and comparison with observations}
\label{sect:results}

The present deflagration simulation of a thermonuclear supernova makes
distinct predictions for observables that can be tested against
real SNe Ia. With certain observables like polarimetry spectra and
spectra of the nebular phase it may be possible to check details of
the results of pure deflagration models against observations and
potentially assess its validity. This certainly warrants further
attention and will be addressed in future work. Here, we are concerned
with some more basic global properties which test whether our
particular simulation falls into the range of observations.

The asymptotic kinetic energy amounted to $8.1 \times
10^{50}\,\mathrm{erg}$. In the hydrodynamical simulation, $0.606\, M_\odot$ of
iron group elements and $0.431\, M_\odot$ of intermediate mass
elements were produced, leaving
behind 0.368 $\, M_\odot$ of unburned carbon and oxygen.

\begin{figure}
\plotone{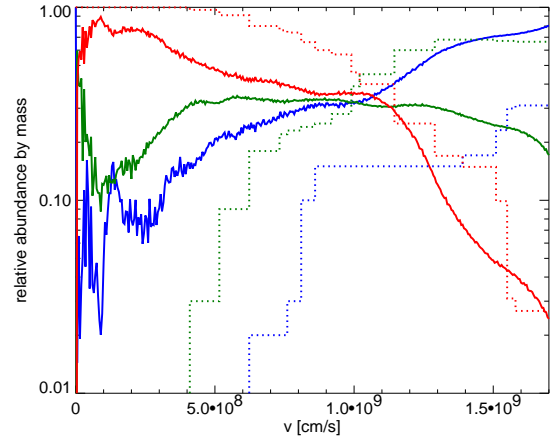}
\caption{Spherically averaged composition resulting from the
  hydrodynamical explosion simulation (solid lines) compared to the
  findings of the abundance tomography of SN~2002bo (dotted lines). Iron group
  element abundances are shown in red, intermediate mass elements in
  green and unburned material in blue. \label{fig:at}}
\end{figure}

The composition of the ejecta can be directly compared with that of observed 
supernovae obtained from abundance tomography \citep{stehle2005a} using a
spherical average over the entire ejecta of the model in homologous expansion. 
Such a comparison with the abundance tomography of SN~2002bo is shown in
Fig.~\ref{fig:at}, where the velocity is used as radial
coordinate. Two cautionary remarks are necessary here. First, SN~2002bo was a
much more energetic and luminous event than that expected to result from
the simulation. Second, as pointed out in Sect.~\ref{sect:evo}, there
exist features with high concentrations of particular species (see
Fig.~\ref{fig:comp3d}) large enough to dominate spectral
observations. The procedure of spherically averaging the mass fractions
washes out these structures and artificially increases the mixing of species.
Nonetheless, the chemical composition of the
ejecta of the simulated explosion and the abundance tomography of
SN~2002bo compare surprisingly well. In both cases, the central parts are
clearly dominated by iron group elements and the overall distribution of the
main species in our simulation generally agrees with that obtained from the
abundance tomography. 
There is good agreement up to velocities of $\sim$$10,000 \, \mathrm{km}\,
\mathrm{s}^{-1}$, if we keep in mind that SN~2002bo had a $^{56}$Ni content
of $\sim 0.5\, M_\odot$, which is significantly larger than in the simulation. 
The slope of the decline of the iron group abundance
towards larger radii is also well-reproduced. These elements are mixed out
to the same radii in the simulation and in the observation. Moreover,
both agree in intermediate
mass elements being more abundant than unburned material below
$\sim$$10,000 \, \mathrm{km}\, \mathrm{s}^{-1}$ and mixed compositions
over a wide range of velocities. The abundance of unburned material in
the simulation is low in this velocity range, in particular near the
center. As previously pointed out by 
\citet{kozma2005a}, the presence of large amounts of unburned fuel at
low velocities gives rise to a pronounced oxygen line in the late-time
("nebular") spectra which is inconsistent with observations. Owing to
the more complete burning of the center with the chosen flame ignition
condition, this problem is alleviated in the current simulation. The
abundance of fuel material near the center is lower by a factor of 10 to 50
than the iron group mass fraction here. 

Still, the abundance
of carbon and oxygen seems rather high in comparison with the
observation. This may in part be an
artifact of the spherical averaging procedure for determining the
composition of the ejecta in the simulation. 
overestimating the contribution of the dense but narrow fuel regions 
(see Fig.~\ref{fig:comp3d}). As the volume of ejecta that influences
the spectra taken form a certain line of sight varies with time, a
careful analysis of this effect would require the derivation of spectra
from our simulation in a three-dimensional approach.

At velocities above $\sim$$10,000 \, \mathrm{km}\,\mathrm{s}^{-1}$,
however, there is a significant discrepancy between the abundance
tomography data of SN 2002bo and the simulated composition. While in the
former the intermediate mass elements dominate out to high velocities,
the ejecta in the simulation quickly become dominated by unburned material. This
discrepancy, however, is only found in the outer $0.25\,M_\odot$.

In order to derive light curves from the simulation, more detailed
information about the chemical composition of the ejecta is necessary
than the data obtained directly from the hydrodynamical explosion
simulation. This information can be derived in a nucleosynthetic
postprocessing step \citep{travaglio2004a} on the basis of
$53^3$ tracer particles that have been 
advected in the flow. The mass of the star was equally distributed
between those tracers and they add a "Lagrangian component" to the
otherwise Eulerian approach. Recording representative temperature and
density profiles they facilitate the \emph{a posteriori}
reconstruction of the nuclear reactions with a large reaction network
(which would be impractical to employ concurrently with the explosion
simulation). This way, the chemical 
composition of the ejecta as needed for the radiative transfer is
determined. In particular, it is necessary to separate the
$^{56}$Ni, which by radioactive decay provides the energy for the
optical event, from the stable iron group isotopes that do not
contribute to the luminosity but are important for absorption. The
postprocessing yields $0.56 \, M_\odot$ of iron group elements ($0.05
\, M_\odot$ less than the estimate from the hydrodynamical simulation)
and $0.33 \, M_\odot$ of $^{56}$Ni. This indicates that the simulated
supernova would range among the lower end of observed normal
SNe~Ia. For events like SN~1991M, SN~1993L, SN~2004eo, and SN~2001el
similar masses of $^{56}$Ni and total iron group elements were
inferred by \citet{mazzali2007a}.

\begin{figure}
\plotone{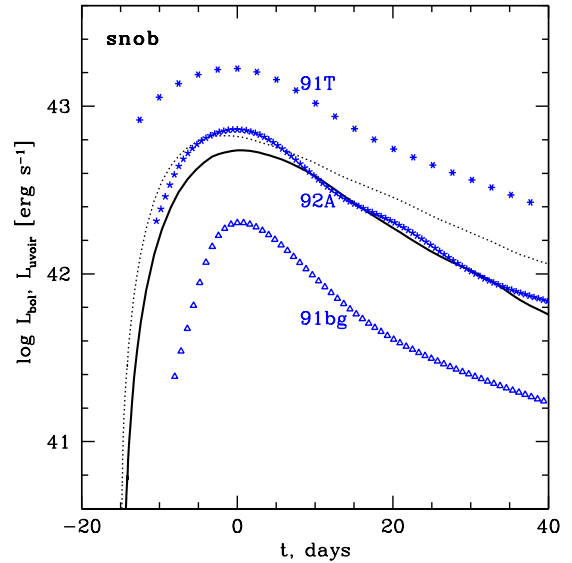}
\caption{Bolometric light curve derived for our model (black curves;
  solid is the ``UVOIR-bolometric'' light curve and the complete
  bolometric light curve is dotted.) The blue dotted curves correspond
  to observed bolometric light curves \citep{stritzinger2006a}. \label{fig:lc_bol}}
\end{figure}

On the basis of the ejecta composition as derived from the
postprocessing step and spherically averaged over the entire star,
bolometric light curves (true bolometric as well as
``UVOIR-bolometric'') have been derived with the radiative transport
code \textsc{Stella}, as described in detail by
\citet{blinnikov2006a}. The result is shown in Fig.~\ref{fig:lc_bol}
and compared to the bolometric light curves of standard SNe Ia as
given by \citet{stritzinger2006a}. Evidently, the synthetic light
curve derived from the simulation is a normal SN Ia light curve both in
terms of the peak luminosity and the shape. It is slightly dimmer than
SN 1992A, as expected from the lower $^{56}$Ni mass
(\citealt{stritzinger2006a} report $0.4\, M_\odot$
of $^{56}$Ni for SN~1992A), but has a very similar rise and decline
rate.

\section{Limits of the Pure Deflagration model}
\label{sect:limits}

The limited success of the presented simulation in reproducing the
observables of SNe Ia raises the question of whether it is possible to
assess generic limits of the pure deflagration model on this basis. 

As a realization in a single
simulation does not cover the full parameter space, this has to be
discussed in combination with simulations performed
previously.
\citet{reinecke2002d,roepke2004c,travaglio2005a,roepke2006b,roepke2006a},
and \citet{schmidt2006a}
explored the impact of various initial conditions on the explosion
process. These simulations were not performed with the same
resolution as the one presented here and are partially based on the
simpler subgrid-scale turbulence model proposed by
\cite{niemeyer1995b}. Still these results can guide an assessment of
the deflagration model since the changes between the two subgrid-scale
models are moderate \citep[at most $~$30\% in the iron group element
production and the energy release;][]{schmidt2006c}.

The two main shortcomings identified in Sect.~\ref{sect:results} were
(i) the overall low explosion strength resulting in a low $^{56}$Ni
production and a low explosion energy and (ii) the underproduction of
intermediate mass elements in the outer layers of the ejecta.

Problem (i) has been noted in all deflagration simulations based on a
consistent approach of modeling the flame propagation. There are,
however, several initial parameters that may remedy the problem. These
have been tested in previous simulations. 

Clearly, a major parameter
is the geometrical configuration of the flame ignition. With simple
central ignitions (``C3''-setups of
\citealp{reinecke2002d,roepke2005b}) the asymptotic kinetic energy of
the ejecta falls around $4.5 \times 10^{50}\, \mathrm{erg}$ and the
masses of produced iron group elements hardly exceed $0.5\,M_\odot$
with $\sim 0.3\, M_\odot$ of it being $^{56}$Ni. This would account
for the dimmest examples among the ``normal'' SNe Ia. Ignition in
multiple spots around the center evidently increases the explosion
strength. For a model ignited in 30 bubbles per octant,
\citet{travaglio2004a} report an asymptotic kinetic energy of about
$6.5 \times 10^{50} \,\mathrm{erg}$ and a $^{56}$Ni production of
$0.418\,M_\odot$. \cite{roepke2006a} explored the impact of multi-spot
ignitions in a systematic approach and found an optimum of such
configurations resulting in an asymptotic kinetic energy of $6.8\times
10^{50}\,\mathrm{erg}$ and $0.67\,M_\odot$ of iron group elements. The
latter value may increase in modeling the ignition as a stochastic
process in time to $\sim 0.75 \, M_\odot$ \citep{schmidt2006a}. Given these values, the
choice of the ignition configuration in the present model was
moderately optimistic producing $0.606\,M_\odot$ of iron
group elements, about $0.33\,M_\odot$ of which is $^{56}$Ni.
A large ratio of produced stable to radioactive iron
group elements (here it amounts to 0.59) seems to be a general feature
of turbulent deflagration 
models ignited in a multitude of ignition kernels since most of the
iron group synthesis proceeds here at the early stages where high
densities prevail such that electron capture reactions are
significant.

Other parameters that have an impact on the $^{56}$Ni production and
the energy release are the central density of the WD at ignition and
its metallicity. A higher central density naturally increases the
amount of material available that potentially can burn to iron group
elements. Moreover, the increased gravitational acceleration will
enhance the buoyancy-induced turbulence and accelerate the flame. On
the other hand, with increased densities electron captures become more
important and will favor the production of stable iron group
isotopes over $^{56}$Ni. The effect of varying the central density has
been addressed by \citet{roepke2006b}. In a density range where
electron captures are still moderate, increasing the central
density by a factor of 2.6 changes the $^{56}$Ni yield only by less
than 10\% although the total iron group element synthesis was enhanced
by $\sim$30\%. Therefore, although energetic results and enhanced production
of the iron group are expected with higher central densities of the
exploding WD star, as electron capture reactions become
important the fraction of $^{56}$Ni in the NSE material and thus the
brightness of the corresponding events will decrease. An
overproduction of stable iron group nuclei is anticipated here.
Turning this argument around, a way to obtain higher $^{56}$Ni masses
(although the total mass of iron group and intermediate mass elements
would not increase) would be to consider a WD with lower central
density ignited in a multi-spot scenario. This issue will be addressed in a forthcoming study, but as
the results of \citet{roepke2006b} indicate the effect is not expected to be
sufficient to account for the brighter observed SN Ia events. 
 
The metallicity of the WD material will have little effect on the iron
group production and the energy release, but evidently affects the
ratio of stable to radioactive iron group element masses. Increasing
the metallicity favors the production of neutron-rich stable nuclei
over $^{56}$Ni \citep{timmes2003a,travaglio2005a,roepke2006a} and
makes the corresponding event dimmer. The $^{56}$Ni masses decrease
linearly with metallicity and the values given above were derived
under the assumption of solar metallicity. Thus, metallicity effects
have no potential of increasing the $^{56}$Ni yields of the
deflagration model significantly.

Issue (ii) is equivalent with the incomplete burning of the outer
layers of the WD star. Here, the problem of modeling the flame
propagation at conditions where the flamelet picture breaks down
becomes important. It has beens shown that a flame acceleration in
this phase would significantly enhance the production of intermediate
mass elements and the energy release \cite{roepke2005d}, but
\citet{schmidt2007a} argues that such an acceleration does not
occur and that a continuation of modeling the flame propagation as in
the flamelet regime is possible. This approach was followed here,
too. Therefore, within the pure turbulent deflagration regime it
seems unrealistic to enhance burning at low densities beyond what was
achieved in the current simulation.

We thus conclude that the initial parameters of the model
are unlikely to leave enough room to adjust the simulations in order
to account for the more powerful SN Ia events and to reproduce the
chemical composition in the outer layers of the ejecta. As the model
of flame propagation itself leaves no free parameters, only such initial
parameters of the WD and the flame ignition come into consideration
here. These have been tested and adjusting them in a favorable way
allows to enter the range of the weaker ``normal'' SNe
Ia in terms of $^{56}$Ni production, explosion energy, and composition
of the inner part of the ejecta. The more powerful observed events,
producing close to $1 \, M_\odot$ of $^{56}$Ni
\citep[e.g.][]{mazzali2007a} are clearly out of reach for the pure
deflagration scenario of SNe Ia.

\section{Conclusions}

We presented a three-dimensional deflagration simulation of a thermonuclear
supernova explosion with the astrophysical modeling and the numerical
implementation designed in a way to allow for self-consistent
simulation. The model contains no tunable parameters except for the 
flame ignition condition, which is physically not well determined yet,
and the initial conditions of the WD which may vary from event to
event in nature. 
An additional uncertainty of the model is due to burning at the lowest
fuel densities which needs 
further attention, but is expected to lead to only minor changes in the
conclusions about the gross properties of the simulation which we present
here. 
Clearly, the most significant parameter is the flame ignition configuration
deciding even about the success of the
deflagration burning to explode the white dwarf star
\cite[e.g.][]{calder2004a,plewa2004a,roepke2007a}. Here, it was set up
in a favorable way in order to burn the fuel material at the center
of the star as completely as possible and to maximize the energy and
$^{56}$Ni production.

The goal of the simulation was to reach a regime in which the
flame/turbulence interaction modeling is fully self-consistent which
requires to resolve the onset of the turbulent cascade on the
computational grid. This was achieved by performing the simulation
with unprecedented resolution. Such a consistent implementation of the
burning model allows the direct confrontation with observables. The
comparison with bolometric light curves and the chemical composition
of the ejecta derived from observed spectra showed that the simulated
explosion indeed reproduces the characteristics
of the observed  fainter examples of normal SNe~Ia. Gross properties
of the simulations such as energy the
energy release and the $^{56}$Ni production were also consistent with
expectations from SN~Ia observations. 
One has to note, however, that even choosing favorable
ignition conditions \citep[a slight improvement is possible with a
time-dependent stochastic distribution of ignition
kernels,][]{schmidt2006a}, the explosion energy and the brightness of
the model event fall towards the lowest values observed for normal SNe
Ia. 

Thus, upon comparison with previous results from lower resolved
simulations exploring the range of the initial parameters, we conclude that
within the 
astrophysical model employed here, it seems impossible to reproduce
the more energetic events of the observed SN~Ia sample. Also, a smooth
shell of intermediate
mass elements surrounding the iron-rich core of the ejecta cloud and
the anti-correlation between the iron-group and the intermediate mass
elements in the ejecta \citep{mazzali2007a} are difficult to explain
within this framework. Therefore an extension of the scenario seems
necessary. The shortcomings of the model in 
reproducing the chemical composition of the ejecta in the outer parts while
providing good agreement with the observations in central regions may
taken as an indication that the deflagration phase is
only the first phase of the explosion mechanism. While the reasonable
agreement with observables of weaker SN~Ia events indicates that it
may be the dominant part of the explosion mechanism here, the
discrepancy is obvious for the brighter parts of the SN~Ia
sample. Possible extensions of the model 
include an improved description of burning in the distributed burning
regime at low fuel densities as well as a transition to a supersonic
detonation mode of flame propagation at the onset of this regime.
Recently, \cite{roepke2007b} modeled delayed detonation scenarios in
three-dimensions under the assumption that the
deflagration-to-detonation transition occurs as the flame enters the
distributed burning regime. Depending on the configuration of the
deflagration flame ignition, a wide range of $^{56}$Ni masses was
recovered which would reproduce observations of normal to bright
SNe~Ia. In this context, the fainter examples correspond to events
with a pronounced deflagration phase. For these, most of the energy
and $^{56}$Ni 
production originates from that phase and the character of the events is
similar to pure deflagration models. Thus, our results may be
interpreted as an indication that the fainter sub-sample of normal
SNe~Ia expose the hallmark of the deflagration phase. Since the
strength of deflagration phase determined the overall characteristics
of the delayed detonation model implemented this way, a careful and
self-consistent treatment of the burning physics in the deflagration
phases is a premise for credible modeling here as well.

\begin{acknowledgements}
The simulation presented here was carried out at the Computer Center
of the Max Planck Society, Garching, Germany. The authors gratefully
acknowledge the hospitality and the inspiring atmosphere at the Kavli
Institute for Theoretical Physics at the University of California
Santa Barbara, where the paper was written. This work was supported by
the European Research Training Network ``The Physics of Type Ia
Supernovae'' under contract HPRN-CT-2002-00303 and the Deutsche
Forschungsgemeinschaft via the Transregional Collaborative Research
Centre TRR 33 ``The Dark Universe''. F.K.R. received support from the NASA
Theory Program (NNG05GG08G) and from the SciDAC program of the DOE
(DE-FC02-01ER41176).
\end{acknowledgements}


\begin{thebibliography}{44}

\bibitem[{{Arnett} {et~al.}(1971){Arnett}, {Truran}, \&
  {Woosley}}]{arnett1971a}
{Arnett}, W.~D., {Truran}, J.~W., \& {Woosley}, S.~E. 1971, \apj, 165, 87

\bibitem[{{Blinnikov} {et~al.}(2006){Blinnikov}, {R{\"o}pke}, {Sorokina},
  {Gieseler}, {Reinecke}, {Travaglio}, {Hillebrandt}, \&
  {Stritzinger}}]{blinnikov2006a}
{Blinnikov}, S.~I., {R{\"o}pke}, F.~K., {Sorokina}, E.~I., {Gieseler}, M.,
  {Reinecke}, M., {Travaglio}, C., {Hillebrandt}, W., \& {Stritzinger}, M.
  2006, \aap, 453, 229

\bibitem[{{Bravo} \& {Garc{\'{\i}}a-Senz}(2006)}]{bravo2006a}
{Bravo}, E. \& {Garc{\'{\i}}a-Senz}, D. 2006, \apjl, 642, L157

\bibitem[{{Cabot} \& {Cook}(2006)}]{cabot2006a}
{Cabot}, W.~H. \& {Cook}, A.~W. 2006, Nature Physics, 2, 562

\bibitem[{{Calder} {et~al.}(2004){Calder}, {Plewa}, {Vladimirova}, {Lamb}, \&
  {Truran}}]{calder2004a}
{Calder}, A.~C., {Plewa}, T., {Vladimirova}, N., {Lamb}, D.~Q., \& {Truran},
  J.~W. 2004, astro-ph/0405126

\bibitem[{{Filippenko}(1997)}]{filippenko1997a}
{Filippenko}, A.~V. 1997, \araa, 35, 309

\bibitem[{{Fryxell} {et~al.}(1989){Fryxell}, {M{\"u}ller}, \&
  {Arnett}}]{fryxell1989a}
{Fryxell}, B.~A., {M{\"u}ller}, E., \& {Arnett}, W.~D. 1989, Hydro\-dynamics
  and nuclear burning, MPA Green Report 449, Max-Planck-Institut f\"ur
  Astrophysik, Garching

\bibitem[{{Gamezo} {et~al.}(2003){Gamezo}, {Khokhlov}, {Oran}, {Chtchelkanova},
  \& {Rosenberg}}]{gamezo2003a}
{Gamezo}, V.~N., {Khokhlov}, A.~M., {Oran}, E.~S., {Chtchelkanova}, A.~Y., \&
  {Rosenberg}, R.~O. 2003, Science, 299, 77

\bibitem[{{Garcia-Senz} \& {Woosley}(1995)}]{garcia1995a}
{Garcia-Senz}, D. \& {Woosley}, S.~E. 1995, \apj, 454, 895

\bibitem[{{Hillebrandt} \& {Niemeyer}(2000)}]{hillebrandt2000a}
{Hillebrandt}, W. \& {Niemeyer}, J.~C. 2000, \araa, 38, 191

\bibitem[{{Hoyle} \& {Fowler}(1960)}]{hoyle1960a}
{Hoyle}, F. \& {Fowler}, W.~A. 1960, \apj, 132, 565

\bibitem[{{Iapichino} {et~al.}(2006){Iapichino}, {Br{\"u}ggen}, {Hillebrandt},
  \& {Niemeyer}}]{iapichino2006a}
{Iapichino}, L., {Br{\"u}ggen}, M., {Hillebrandt}, W., \& {Niemeyer}, J.~C.
  2006, \aap, 450, 655

\bibitem[{{Kozma} {et~al.}(2005){Kozma}, {Fransson}, {Hillebrandt},
  {Travaglio}, {Sollerman}, {Reinecke}, {R{\"o}pke}, \&
  {Spyromilio}}]{kozma2005a}
{Kozma}, C., {Fransson}, C., {Hillebrandt}, W., {Travaglio}, C., {Sollerman},
  J., {Reinecke}, M., {R{\"o}pke}, F.~K., \& {Spyromilio}, J. 2005, \aap, 437,
  983

\bibitem[{{Mazzali} {et~al.}(2007){Mazzali}, {R{\"o}pke}, {Benetti}, \&
  {Hillebrandt}}]{mazzali2007a}
{Mazzali}, P.~A., {R{\"o}pke}, F.~K., {Benetti}, S., \& {Hillebrandt}, W. 2007,
  Science, 315, 825

\bibitem[{{Niemeyer} \& {Hillebrandt}(1995)}]{niemeyer1995b}
{Niemeyer}, J.~C. \& {Hillebrandt}, W. 1995, \apj, 452, 769

\bibitem[{{Nomoto} {et~al.}(1976){Nomoto}, {Sugimoto}, \& {Neo}}]{nomoto1976a}
{Nomoto}, K., {Sugimoto}, D., \& {Neo}, S. 1976, \apss, 39, L37

\bibitem[{{Osher} \& {Sethian}(1988)}]{osher1988a}
{Osher}, S. \& {Sethian}, J.~A. 1988, J. Comp. Phys., 79, 12

\bibitem[{{Peters}(2000)}]{peters2000a}
{Peters}, N. 2000, Turbulent Combustion (Cambridge: Cambridge University Press)

\bibitem[{{Plewa} {et~al.}(2004){Plewa}, {Calder}, \& {Lamb}}]{plewa2004a}
{Plewa}, T., {Calder}, A.~C., \& {Lamb}, D.~Q. 2004, \apjl, 612, L37

\bibitem[{{Reinecke} {et~al.}(1999{\natexlab{a}}){Reinecke}, {Hillebrandt}, \&
  {Niemeyer}}]{reinecke1999b}
{Reinecke}, M., {Hillebrandt}, W., \& {Niemeyer}, J.~C. 1999{\natexlab{a}},
  \aap, 347, 739

\bibitem[{{Reinecke} {et~al.}(2002{\natexlab{a}}){Reinecke}, {Hillebrandt}, \&
  {Niemeyer}}]{reinecke2002b}
---. 2002{\natexlab{a}}, \aap, 386, 936

\bibitem[{{Reinecke} {et~al.}(2002{\natexlab{b}}){Reinecke}, {Hillebrandt}, \&
  {Niemeyer}}]{reinecke2002d}
---. 2002{\natexlab{b}}, \aap, 391, 1167

\bibitem[{{Reinecke} {et~al.}(1999{\natexlab{b}}){Reinecke}, {Hillebrandt},
  {Niemeyer}, {Klein}, \& {Gr{\" o}bl}}]{reinecke1999a}
{Reinecke}, M., {Hillebrandt}, W., {Niemeyer}, J.~C., {Klein}, R., \& {Gr{\"
  o}bl}, A. 1999{\natexlab{b}}, \aap, 347, 724

\bibitem[{{R{\"o}pke}(2005)}]{roepke2005c}
{R{\"o}pke}, F.~K. 2005, \aap, 432, 969

\bibitem[{{R{\"o}pke} {et~al.}(2006{\natexlab{a}}){R{\"o}pke}, {Gieseler},
  {Reinecke}, {Travaglio}, \& {Hillebrandt}}]{roepke2006b}
{R{\"o}pke}, F.~K., {Gieseler}, M., {Reinecke}, M., {Travaglio}, C., \&
  {Hillebrandt}, W. 2006{\natexlab{a}}, \aap, 453, 203

\bibitem[{{R{\"o}pke} \& {Hillebrandt}(2004)}]{roepke2004c}
{R{\"o}pke}, F.~K. \& {Hillebrandt}, W. 2004, \aap, 420, L1

\bibitem[{{R{\"o}pke} \& {Hillebrandt}(2005{\natexlab{a}})}]{roepke2005b}
---. 2005, \aap, 431, 635

\bibitem[{{R{\"o}pke} \& {Hillebrandt}(2005{\natexlab{b}})}]{roepke2005d}
---. 2005, \aap, 429, L629

\bibitem[{{R{\"o}pke} {et~al.}(2006{\natexlab{b}}){R{\"o}pke}, {Hillebrandt},
  {Niemeyer}, \& {Woosley}}]{roepke2006a}
{R{\"o}pke}, F.~K., {Hillebrandt}, W., {Niemeyer}, J.~C., \& {Woosley}, S.~E.
  2006{\natexlab{b}}, \aap, 448, 1

\bibitem[{{R{\"o}pke} \& {Niemeyer}(2007)}]{roepke2007b}
{R{\"o}pke}, F.~K. \& {Niemeyer}, J.~C. 2007, \aap, 464, 683

\bibitem[{{R{\"o}pke} {et~al.}(2007){R{\"o}pke}, {Woosley}, \&
  {Hillebrandt}}]{roepke2007a}
{R{\"o}pke}, F.~K., {Woosley}, S.~E., \& {Hillebrandt}, W. 2007, \apj, 660,
  1344

\bibitem[{{Ruiz-Lapuente} {et~al.}(2004){Ruiz-Lapuente}, {Comeron},
  {M{\'e}ndez}, {Canal}, {Smartt}, {Filippenko}, {Kurucz}, {Chornock}, {Foley},
  {Stanishev}, \& {Ibata}}]{ruiz-lapuente2004a}
{Ruiz-Lapuente}, P., {Comeron}, F., {M{\'e}ndez}, J., {Canal}, R., {Smartt},
  S.~J., {Filippenko}, A.~V., {Kurucz}, R.~L., {Chornock}, R., {Foley}, R.~J.,
  {Stanishev}, V., \& {Ibata}, R. 2004, \nat, 431, 1069

\bibitem[{{Schmidt}(2007)}]{schmidt2007a}
{Schmidt}, W. 2007, \aap, 465, 263

\bibitem[{{Schmidt} \& {Niemeyer}(2006)}]{schmidt2006a}
{Schmidt}, W. \& {Niemeyer}, J.~C. 2006, \aap, 446, 627

\bibitem[{{Schmidt} {et~al.}(2006{\natexlab{a}}){Schmidt}, {Niemeyer}, \&
  {Hillebrandt}}]{schmidt2006b}
{Schmidt}, W., {Niemeyer}, J.~C., \& {Hillebrandt}, W. 2006{\natexlab{a}},
  \aap, 450, 265

\bibitem[{{Schmidt} {et~al.}(2006{\natexlab{b}}){Schmidt}, {Niemeyer},
  {Hillebrandt}, \& {R{\"o}pke}}]{schmidt2006c}
{Schmidt}, W., {Niemeyer}, J.~C., {Hillebrandt}, W., \& {R{\"o}pke}, F.~K.
  2006{\natexlab{b}}, \aap, 450, 283

\bibitem[{{Smiljanovski} {et~al.}(1997){Smiljanovski}, {Moser}, \&
  {Klein}}]{smiljanovski1997a}
{Smiljanovski}, V., {Moser}, V., \& {Klein}, R. 1997, Combustion Theory and
  Modelling, 1, 183

\bibitem[{{Stehle} {et~al.}(2005){Stehle}, {Mazzali}, {Benetti}, \&
  {Hillebrandt}}]{stehle2005a}
{Stehle}, M., {Mazzali}, P.~A., {Benetti}, S., \& {Hillebrandt}, W. 2005,
  \mnras, 360, 1231

\bibitem[{{Stritzinger} {et~al.}(2006){Stritzinger}, {Leibundgut}, {Walch}, \&
  {Contardo}}]{stritzinger2006a}
{Stritzinger}, M., {Leibundgut}, B., {Walch}, S., \& {Contardo}, G. 2006, \aap,
  450, 241

\bibitem[{{Timmes} {et~al.}(2003){Timmes}, {Brown}, \& {Truran}}]{timmes2003a}
{Timmes}, F.~X., {Brown}, E.~F., \& {Truran}, J.~W. 2003, \apjl, 590, L83

\bibitem[{{Travaglio} {et~al.}(2005){Travaglio}, {Hillebrandt}, \&
  {Reinecke}}]{travaglio2005a}
{Travaglio}, C., {Hillebrandt}, W., \& {Reinecke}, M. 2005, \aap, 443, 1007

\bibitem[{{Travaglio} {et~al.}(2004){Travaglio}, {Hillebrandt}, {Reinecke}, \&
  {Thielemann}}]{travaglio2004a}
{Travaglio}, C., {Hillebrandt}, W., {Reinecke}, M., \& {Thielemann}, F.-K.
  2004, \aap, 425, 1029

\bibitem[{{Wang} {et~al.}(2007){Wang}, {Baade}, \& {Patat}}]{wang2007a}
{Wang}, L., {Baade}, D., \& {Patat}, F. 2007, Science, 315, 212

\bibitem[{{Woosley} {et~al.}(2004){Woosley}, {Wunsch}, \&
  {Kuhlen}}]{woosley2004a}
{Woosley}, S.~E., {Wunsch}, S., \& {Kuhlen}, M. 2004, \apj, 607, 921

\bibitem[{{Zingale} {et~al.}(2005){Zingale}, {Woosley}, {Rendleman}, {Day}, \&
  {Bell}}]{zingale2005a}
{Zingale}, M., {Woosley}, S.~E., {Rendleman}, C.~A., {Day}, M.~S., \& {Bell},
  J.~B. 2005, \apj, 632, 1021

\end{thebibliography}
\end{document}